\documentclass[preprint,showpacs,amsmath,amssymb]{revtex4}

\usepackage{graphicx}
\usepackage{dcolumn}
\usepackage{bm}

\def\ddpopa#1#2{{{d#1}\over{d#2}}}

\def\be{\begin{equation}}
\def\ee{\end{equation}}

\def\B.#1{{\it {\bf #1}}}

\def\be{\begin{equation}}
\def\ee{\end{equation}}
\def\bse{\begin{subequations}}
\def\ese{\end{subequations}}

\def\etal{{\em et al.}}

\def\2d{two-dimensional }
\def\3d{three-dimensional }
\def\ns{Navier-Stokes }
\def\r{Reynolds number}
\def\C.#1{{\cal{#1}}}

\def \beq {\begin{equation}}
\def \eeq {\end{equation}}
\def \beqa {\begin{eqnarray}}
\def \eeqa {\end{eqnarray}}

\def \minp {\begin{minipage}{0.85\textwidth}}
\def \mine {\end{minipage}}
\def \minpfig {\begin{minipage}{0.93\textwidth}}
\def \minefig{\end{minipage}}

\def\nsig2{n_2 \sigma_2}
\def\ns1{n_1 \sigma_1}
\def\tbf{\textbf}

\begin{document}


\title{Preventing transition to turbulence: a viscosity stratification does not 
always help}
\author{Vijayakumar Chikkadi}
\author{A. Sameen}
\author{Rama Govindarajan}%
 \email{rama@jncasr.ac.in}
\affiliation{Engineering Mechanics Unit,
Jawaharlal Nehru Centre for Advanced Scientific Research, Jakkur,
Bangalore 560064, India \\
}

\date{\today}

\begin{abstract}
In channel flows a step on the route to turbulence
is the formation of streaks, often due to algebraic growth of disturbances.
While a variation of viscosity in the gradient direction often plays a large
role in laminar-turbulent transition in shear flows, we show that it has,
surprisingly, little effect on the algebraic growth.
Non-uniform viscosity therefore may not always work as a flow-control 
strategy for maintaining the flow as laminar.
\end{abstract}

\maketitle

In many flow applications, preventing laminar flow from undergoing
a transition to turbulence is highly desirable. One of the most popular 
methods has been to employ a stratification of viscosity in the direction
normal to the wall. A reduction of near-wall viscosity by the addition of
shear-thinning substances, or by heating/ 
cooling the walls, can have a large stabilising effect on the {\em linear} 
disturbance modes in a laminar flow \cite{yih,wall,pina2,rama2}. However,
in internal shear flows such as those through pipes and channels 
\cite{holg03,hof03}, the flow becomes turbulent while it is still linearly
stable. For example, the laminar flow through a two-dimensional 
channel is linearly stable upto a critical Reynolds number $Re_{cr}$, based on 
channel half-width and centerline velocity, of $5772$. If the channel as well 
as the incoming flow are specially designed to be extremely quiet, the 
flow can be kept laminar beyond the critical {\r} for linear instability, 
as Nishioka \etal \cite{nish} demonstrated. However, in a typical channel, 
in the absence of such special care, transition to turbulence occurs 
at $Re\sim1000$. The formation of streamwise 
streaks is usually one of the first steps towards turbulence \cite{kling,hof}.
At low levels of external disturbance, these streaks initially form due to 
the algebraic growth of disturbances \cite{hennipof, wal2, Butler, schmid}.
A nonlinear process then makes it possible for the streaks to sustain
themselves. 
(At higher noise levels streak formation could itself be nonlinear.) 
If we are to control the flow, i.e., keep it laminar, by imposing a 
viscosity stratification, we must first know what it does to the initial 
departure from a laminar flow towards a turbulent state, i.e. to streak
formation. Rather unexpectedly, we find here that it has very little 
effect on the algebraic growth mechanism. 
The effect on nonlinear processes in transition, such as the
self-sustenance of streaks needs to be investigated in future.
On the other hand, fully developed turbulence, especially the problem
of drag reduction due to polymers, has been investigated by many.

We study the symmetric flow through a channel of (i) shear-thinning 
fluids with negligible visco-elasticity, such as dispersions, concentrated
colloidal suspensions and carboxymethyl cellulose, and (ii) two miscible
fluids of equal densities but different viscosities.
Here, fluid 1 flows in the region $-p<y<p$ where the walls are at $y= \pm 1$.
Fluid 2 flows in $|y| \ge p+q$, with a thin mixed layer of thickness $q$
between the two where viscosity varies from $\mu_1$ at $p$ to $m \mu_1$ at 
$p+q$. For the purpose of isolating the effect of viscosity variation, and to
make a better comparison with case (i), we have not included the effect of
diffusivity in case (ii).
We note that this work does not give a firm answer on the complex
effect of polymers, but we are able to state firmly that a stratification
of viscosity alone does not affect transient growth. Incidentally, with
the exception of \cite{alison}, who had a completely different
objective, there is no computation to our knowledge of transient growth 
with varying viscosity.

\begin{figure}
\includegraphics[width=0.35\textwidth]{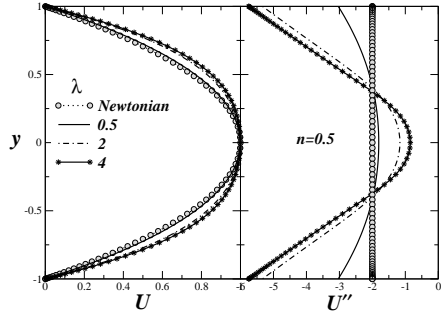}
\caption{Velocity (left) and its second derivative (right) profiles at $n=0.5$ and 
different values of $\lambda$.}
\label{vel0}
\end{figure}

\begin{figure}
\includegraphics[width=0.35\textwidth]{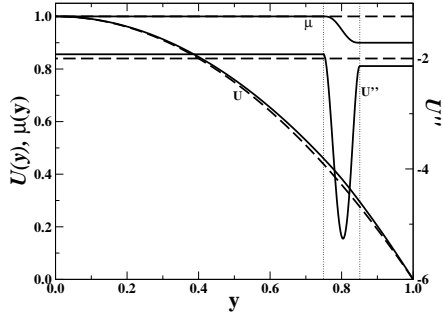}
\caption{The velocity and the viscosity profiles for the two-fluid case 
\cite{rama1}. 
The second derivative of velocity is also shown. Solid lines: two fluids,
$m=0.9$; dashed lines: single fluid. The vertical lines show the extent 
of the mixed region.}
\label{veltf}
\end{figure}

The stability operator is non-orthogonal, which enables linearly stable
eigenmodes to
grow algebraically to give high levels of transient growth. If no
other process intervened, these would eventually decay, but 
nonlinearity takes over when sufficient amplitudes are attained.
We begin by obtaining linear eigenmodes, as already done for case
(ii) in \cite{rama2}. Case (i) is described below, and affords no surprise,
i.e., a shear-thinning viscosity stabilises linear perturbations.
The basic flow velocity is $u=U(y), v=w=0$ 
in the streamwise ($x$), normal to the wall ($y$) and spanwise
($z$) directions respectively. The apparent viscosity of shear-thinning
fluids is a function of the scalar invariants of the shear rate 
$\dot\gamma$. The Carreau model \cite{bird},
\begin{equation}
\frac{\mu-\mu_{\infty}}{\mu_{0}-\mu_{\infty}}=[1+(\lambda\dot\gamma)^{2}]
^{\frac{(n-1)}{2}}
\end{equation}
where $\mu_{0}$ and $\mu_{\infty}$ are the viscosities at zero and 
infinite shear rate respectively, $\lambda$ is the time constant of the 
fluid and $n$ is the shear-thinning index, is known to be a good
representation of the viscosity, $n=1$ or $\lambda=0$ correspond to a 
Newtonian fluid. 
The mean velocity profile is obtained from the steady $x$-momentum 
equation, given in non-dimensional form by
\begin{equation}
-P+\frac{d}{dy}(\mu\frac{dU}{dy})=0
\quad {\rm where } \quad
P \equiv Re\frac{dp}{dx}
\end{equation}
The primes denote differentiation with respect to $y$, and 
$\mu$ is scaled by $\mu_{0}$. The Reynolds number is based on viscosity 
averaged across the channel. Fig. \ref{vel0} shows the velocity and  
its second derivatives with respect to $y$. 
Sample basic profiles for the two-fluid case
are given in Fig. \ref{veltf}, details are available in \cite{rama1}. 

Three-dimensional linear perturbations in the velocity, in normal mode 
form, e.g.,
\beqa
(\hat v,\hat{\eta})(x,y,z,t)=(v,\eta)(y) \hspace{2mm}\exp\{i(\alpha x+
\beta z-\omega t)\},
\eeqa
where $\alpha$ and $\beta$ are their streamwise and spanwise wavenumbers 
respectively and $\omega$ is their frequency,
satisfy the Orr-Sommerfeld and Squire's equations \cite{schmid}, 
which are modified here to include the effects of non-constant viscosity.
The result is an eigenvalue problem described by
\beq
  i\omega\left(
    \begin{array}{cc}
  {k^2-D^2}  & 0  \\
       0 & 1
    \end{array}
  \right)
\left(
    \begin{array}{c}
      v  \\
      \eta
    \end{array}
  \right)=
\left(
    \begin{array}{cc}
      \mathcal{L}_{os}  & 0  \\
    i\beta DU & \mathcal{L}_{sq}
    \end{array}
  \right)
\left(
    \begin{array}{c}
      v  \\
      \eta
    \end{array}
  \right)
\label{oper_eqn}
\eeq
where the modified Orr-Sommerfeld and Squire operators are given
respectively by
\beqa
&&\mathcal{L}_{os}=i\alpha [U (k^2-D^2)+U^{\prime\prime}]
+\frac{1}{Re}[\mu (k^2-D^2)^2+ \nonumber\\
&& 2\mu^{\prime}D^3+\mu^{\prime\prime}D^2-2k^2\mu^{\prime}D+k^2\mu^{\prime\prime}],
\label{los}
\eeqa
\beqa
 \mathcal{L}_{sq}=i\alpha U+\frac{1}{Re}\Big[\mu (k^2
-D^2)+\mu^{\prime} D\Big],
\label{lsq}
\eeqa
\beqa
k^2=\alpha^2+\beta^2,
\eeqa
$D\equiv {d}/{dy}$ and $\eta$ is the normal vorticity of the disturbance.
Equation (\ref{oper_eqn}) along with boundary conditions $v=Dv=\eta=0$ 
at $y=\pm 1$ is solved using a Chebyshev collocation spectral method. 
The number of collocation points was taken to be 81. 
On using 161 collocation points, the growth rates changed only in the 
sixth significant decimal place in the worst case. The difference in
individual eigenvalues was much smaller.
For specified $Re$, $\alpha$ and $\beta$, the
imaginary part of the frequency gives the exponential decay rate.

\begin{figure}
\includegraphics[width=0.35\textwidth]{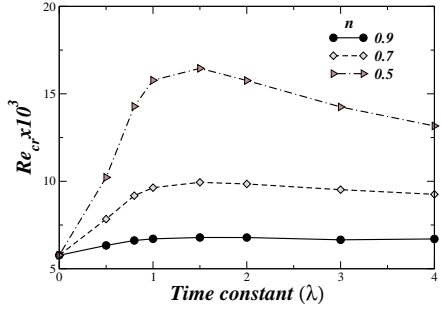}
\caption{Variation of the critical Reynolds number of a shear-thinning
fluid, with $\lambda$, at $n$=0.5, 0.7 and 0.9, $\beta=0$. }
\label{prim1}
\end{figure}

The effect of shear-thinning viscosity on the linear instability is 
shown in Fig. \ref{prim1}, it is evident that shear-thinning stabilises 
the flow. The increase in the critical Reynolds number for instability 
($Re_{cr}$) can be a factor of $3$ over that for a Newtonian fluid. 
For inviscid flow \cite{drazin81} a necessary 
and sufficient condition for flow instability is the existence of an
inflexion point in the mean velocity profile. In a finite Reynolds number
flow no such theorem exists, but it is normally the case that a flow 
stabilisation results when the velocity profile becomes fuller, i.e.,
goes further away from containing a point of inflexion. The second
derivatives of the velocity in Fig. \ref{vel0} indicate that the observed 
effect is the expected one. 
The effect on linear stability was seen \cite{rama2,rama1} to be much more 
dramatic
in the case of two-fluid flow, when the viscosity-stratified layer
overlapped the production layer of disturbance kinetic energy. 
The critical 
Reynolds number can be higher by an order of magnitude. In the
light of this, it is surprising that the dominant algebraic growth of 
disturbances is unmoved by the viscosity stratification. 

\begin{figure}
\includegraphics[width=0.35\textwidth]{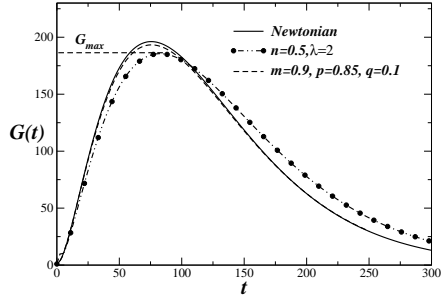}
\caption{Energy amplification $G(t)$ of Newtonian (solid line), shear-thinning fluid (filled circle) and 
two-fluid flow (dash line) at 
$Re=1000,\:\alpha=0.0,\:\beta=2.05$}
\label{gmax_fig}
\end{figure}

\begin{figure}
\includegraphics[width=0.35\textwidth]{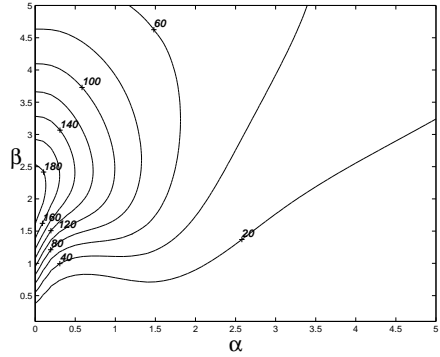}
\caption{Contours of the maximum over time of optimal disturbance 
$G_{max}$ for two-fluid flow. The viscosity ratio is $m$=0.9, $p=0.85$, 
$q=0.1$ and $Re$=1000.
The number mentioned on the contours is the value of $G_{max}$. The figure is 
very similar to that for a single Newtonian fluid in \cite{schmid}.}
\label{gmax09}
\end{figure}

\begin{figure}
\centering
\begin{minipage}{0.4\textwidth}
\includegraphics[height=0.35\textwidth]{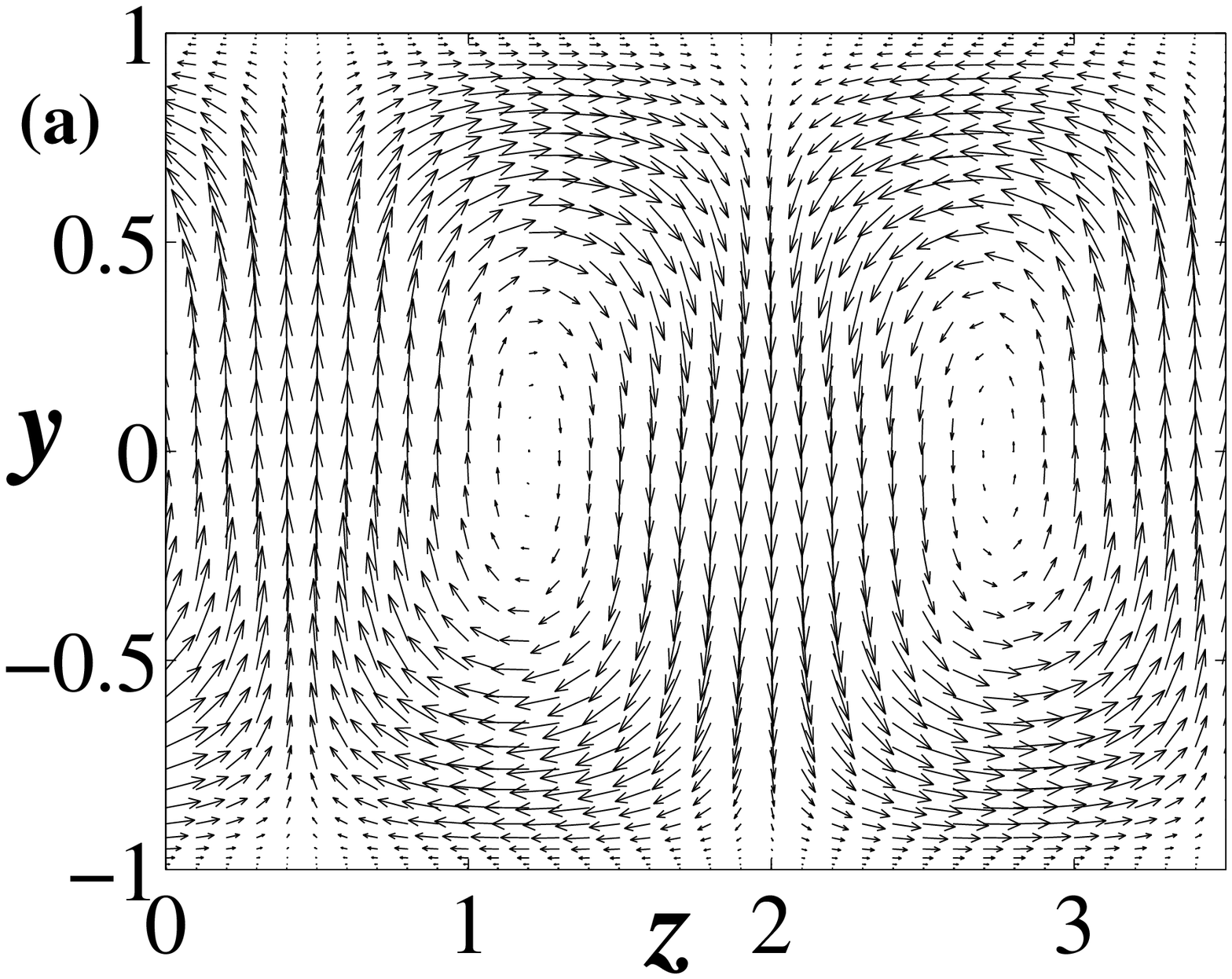}
\includegraphics[height=0.35\textwidth]{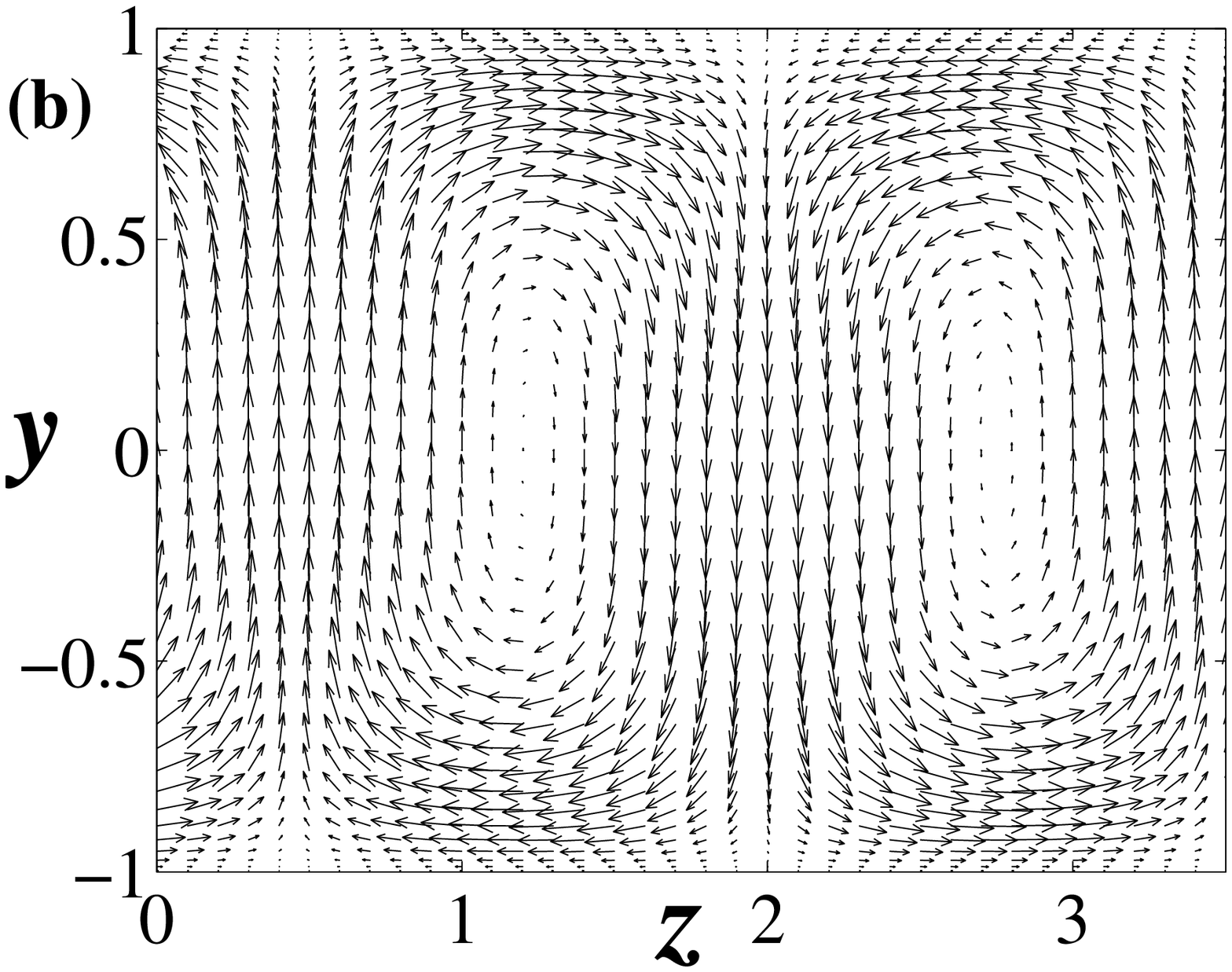}
\end{minipage}
\caption{Optimal disturbances at $\alpha=0.0$, $\beta=2.05$, $Re=1000$. 
(a) Newtonian fluid; 
(b) Shear-thinning fluid, $n=0.5$, $\lambda=2.0$.}
\label{opt_rolls}
\end{figure}

\begin{figure}
\includegraphics[width=0.35\textwidth]{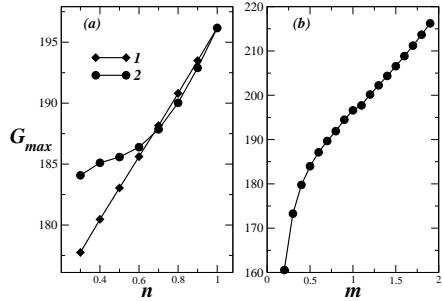}
\caption{Growth of streamwise-independent ($\alpha$=0) disturbances 
at $Re$=1000, $\beta$=2.05. (a) Non-Newtonian fluid, with different levels of 
shear-thinning. (b) Two-fluid flow, $m$ is the ratio
of outer to inner fluid viscosity. $G_{max}$
for a single Newtonian fluid is $\sim 196$.}
\label{opt_dis}
\end{figure}

For studying transient growth,
the disturbance equation (\ref{oper_eqn}) is viewed as an initial value 
problem:
\begin{equation}
\frac{\partial}{\partial t}
\left[
    \begin{array}{c}
      v  \\
      \eta
    \end{array}
  \right]
=\tbf M^{-1}\tbf L
\left[
    \begin{array}{c}
      v  \\
      \eta
    \end{array}
  \right]
\label{ivp}
\end{equation}
\begin{equation}
\tbf M=\left[\begin{array}{cc}
-D^{2}+k^{2} & 0 \\
0 & 1 \\
\end{array}
\right]\hspace{3mm}
\tbf L=\left[\begin{array}{cc}
\mathcal{L}_{os} & 0\\
-i\beta U^{\prime} & \mathcal{L}_{sq}\\
\end{array}
\right]
\end{equation}
where $\mathcal{L}_{os}$ and $\mathcal{L}_{sq}$ are defined by equations
(\ref{los}) \& (\ref{lsq}). We choose to monitor the growth of the disturbance
kinetic energy, given by
\beqa
E_t= \frac{1}{2}
\int^{+1}_{-1}{(|Dv|^2+k^2|v|^2+|\eta|^2)}dy.
\label{energy}
\eeqa
The total kinetic energy of the perturbation can be found by integrating 
equation (\ref{energy}) at a given time over the $\alpha-\beta$ plane.
The amplitude of disturbance kinetic energy at a given time depends,
of course, on the initial amplitudes of various modes. Here, $G$ is
scaled by its initial value. For a given viscosity stratification,
equation (\ref{gmax}) gives the maximum 
possible amplitude $G$ at any instant of time, optimised for each 
instant of time, over all initial conditions:
\begin{equation}
G(t|\alpha,\beta,R)\equiv G(t)= \max \frac{E(t|\alpha,\beta,R)}
{E(0,\alpha,\beta,R)}.
\label{gmax}
\end{equation}
The optimisation method followed here is outlined in \cite{schmid}. 
The maximum of $G$ over time is denoted as $G_{max}$, which is shown in
Fig. \ref{gmax_fig}. It is seen that a variation of viscosity 
does not make much difference to the transient growth of disturbances of
this wavelength.
The maximum energy contour for two-fluid flow, $m=0.9$, is shown in 
Fig. \ref{gmax09}. The corresponding contour for a single Newtonian 
fluid is available in \cite{schmid}. There is very little change anywhere
in the $\alpha-\beta$ due to viscosity variation. A similar conclusion
was reached about the effect of shear-thinning.
It is known 
\cite{Butler,schmid} that in the channel flow of a Newtonian fluid, 
a streamwise vortex, with $\alpha=0$ and $\beta \simeq 2$, is the 
optimal disturbance, i.e., gives the highest $G_{max}$. As seen
in Fig. \ref{gmax09}, we find the same 
to be true of viscosity-stratified flow. 
The optimal disturbance is in the form of rolls, as shown in Fig. 
\ref{opt_rolls}. We now choose these disturbances, that remain
constant in the streamwise direction, and quantify the effect of a varying
viscosity in Fig. \ref{opt_dis}. We see that there is at most a $15\%$ change, 
for cases where the linear stability changed by a factor of $3$ 
(shear-thinning fluid) and by an order of magnitude (two-fluid).
In particular,
when $m=0.9$ in (b), the change is only about $1\%$, whereas linear 
stability changes by an order of magnitude.
It is apparent that 
viscosity stratification is not effective in suppressing the 
growth of optimal disturbances. 

This result is counter-intuitive, 
especially given the major stabilisation achieved by the linear modes,
and a pre-conditioning to expect stabilisation from a fuller velocity profile.
Our attempts to provide a complete mathematical explanation have not been
successful so far. We offer a partial explanation based on the
observation that the largest growing transients are those that do not vary in
the downstream direction, i.e., modes with $\alpha=0$. Out of the terms
containing derivatives of the viscosity in equation (\ref{u2p}) it can be
verified numerically that the only term which contributes any noticeable effect
is that containing $U^{\prime\prime}$. Note that $U^{\prime\prime}$ may be
written as
\be
U^{\prime\prime}=-\frac{P}{\mu}+\frac{Py}{\mu^2}\ddpopa{\mu}{y}.
\label{u2p}
\ee
At $\alpha=0$
the term containing $U^{\prime\prime}$ does not appear in equation 
(\ref{los}), and so the major 
effect of viscosity stratification is absent in the case of growth of
optimal disturbances. The reduced problem has been solved, and it has been
verified that the eigenvalues and eigenfunctions are practically unchanged by
viscosity stratification. This fact is evident from Fig. \ref{opt_rolls} as well.

A viscosity reduction near the wall is usually associated with stabilisation
and with drag reduction. We conclude however, that while the linear modes 
are indeed significantly stabilised, the transient growth of disturbances, 
including that of streamwise vortices, is practically unaffected. Flow 
control using a stratification of viscosity is unlikely to work in this 
case, at least in suppressing the first deviation from a laminar state. 

{\bf Acknowledgements:} We thank Prof. Dan Henningson for discussions.

\end{document}